\documentclass[preprint,aps,floats,floatfix,nofootinbib,superscriptaddress]{revtex4-1}
\usepackage{eurosym}
\usepackage{graphicx,array,dcolumn}
\usepackage{blindtext}
\usepackage{calc,tabularx, epsfig}
\usepackage{hyperref}
\usepackage{amsmath}
\usepackage{amssymb}
\usepackage{multirow}
\usepackage{xspace}
\usepackage{slashed}

\allowdisplaybreaks[1]
\newcolumntype{C}[1]{>{\centering\let\newline\\\arraybackslash\hspace{0pt}}m{#1}}
\newlength{\figurewidth}
\newcommand{\beq}{\begin{equation}}
\newcommand{\eeq}{\end{equation}}
\newcommand{\bea}{\begin{eqnarray}}
\newcommand{\eea}{\end{eqnarray}}
\newcommand{\ba}{\begin{array}}
\newcommand{\ea}{\end{array}}

\makeatother
\begin{document}

\title{AdS backgrounds and induced gravity}
\author{Hai Lin}
\affiliation{Yau Mathematical Sciences Center,
Tsinghua University, Beijing 100084, China.}
\author{Gaurav Narain}
\affiliation{Institute of Theoretical Physics (ITP),
Chinese Academy of Sciences (CAS), Beijing 100190, China.}

\begin{abstract}
In this paper we look for AdS solutions to generalised gravity theories in
the bulk in various spacetime dimensions. The bulk gravity action includes
the action of a non-minimally coupled scalar field with gravity, and a
higher-derivative action of gravity. The usual Einstein-Hilbert gravity is
induced when the scalar acquires a non-zero vacuum expectation value. The
equation of motion in the bulk shows scenarios where AdS geometry emerges
on-shell. We further obtain the action of the fluctuation fields on the
background at quadratic and cubic orders.
\end{abstract}

\maketitle
\vspace{2cm}

\vspace{2cm}

\vspace{3cm}

\vspace{3cm}

\vspace{6cm}

\vspace{6cm}



%

%
%


\section{Introduction}

\label{intro} 

When well known framework of quantum field theory (QFT) on flat spacetime
was applied to Einstein-Hilbert gravity, it was seen that the resulting
quantum theory had no well defined high-energy behavior. In the sense that
the theory was plagued with ultraviolet divergences arising at each loop
order which cannot be absorbed in previous existing terms, thereby making
the theory non-renormalizable \cite{tHooft:1974toh, Goroff:1985sz,
Goroff:1985th}. Moreover the gravitational coupling strength was seen to
grow without a bound. To overcome these hurdles, the idea of asymptotic
safety scenario was suggested which generalizes the notion of
renormalizability \cite{Weinberg:1976xy, Weinberg:1980gg} (see also \cite%
{Percacci:2007sz} for a review). Around the same time another proposal came
into existence: considering gravity theory at high energies by incorporating
higher-derivative terms \cite{Stelle:1976gc}, which was shown to be
perturbatively renormalizable to all loops. Although the theory was seen to
have problems with tree level unitarity, these problems were conjectured to
vanish in quantum theory \cite{Salam:1978fd, Julve:1978xn}. Indeed, it was
seen that under quantum corrections this modified theory evades the problems
of higher-derivative ghosts \cite{Narain:2011gs, Narain:2012nf}. These
theories have been thoroughly investigated with or without matter, at
one-loop \cite{Fradkin:1981hx, Fradkin:1981iu, Alvarez-Gaume:2015rwa,
Barth:1983hb, Avramidi:1985ki, Buchbinder:1989jd, Shapiro:1989dq,
Odintsov:1989gz}.

An interesting proposal born of induced gravity, where the quantum
fluctuation of matter generates dynamics of gravity \cite{Zee:1978wi,
Zee:1980sj, Adler:1980pg, Adler:1980bx, Adler:1980md, Adler:1982ri} with an
Einstein-Hilbert term, was very actively investigated. The idea actually was
proposed earlier \cite{Zeldovich1:967gd, Sakharov:1967pk, Chudnovsky:1976zj}
but later works of Adler and Zee made it more formal. Furthermore, it was
realized that Einstein-Hilbert gravity could be generated from a
Weyl-invariant theory of matter and gravity \cite{Nepomechie:1983yq,
Zee:1983mj}. Recently this proposal of induced gravity coupled with
higher-derivative gravity has gained momentum \cite{Salvio:2016vxi,
Einhorn:2014gfa, Salvio:2014soa, Jones:2015son, Einhorn:2015lzy,
Einhorn:2016mws}, where low energy Einstein-Hilbert gravity is generated via
symmetry breaking in UV well defined quantum theory of scale-invariant
system. This new proposal has wide applications in cosmology too. In fact a
lot of work has already been done in this direction \cite%
{Cooper:1982du,Finelli:2007wb}.

In this paper we consider AdS solutions in the generalised gravity theory,
in particular the induced gravity. In the bulk of the gravity, there is a
non-minimal coupling of the scalar field to the Ricci curvature. The action
we consider is from induced gravity action, where one obtains the usual
Einstein-Hilbert gravity, when the scalar field acquires a non-zero vacuum
expectation value. For a review of induced gravity, see also \cite%
{Adler:1982ri}. We also add a curvature-squared term. The quadratic
curvature-squared terms can appear in low energy effective action of string
compactifications, see for example \cite{Alvarez-Gaume:2015rwa}. The
curvature-squared terms can also appear in brane world-volume actions, and
are relevant in brane-world models. Hence it is important to study the
behavior of such actions.

We find AdS solutions from this action. AdS solutions are particularly
interesting for the AdS/CFT correspondence. This correspondence \cite%
{Maldacena:1997re,Gubser:1998bc,Witten:1998qj} has provided a remarkable way
to study quantum gravity by quantum field theory on the boundary of the
spacetime. It reveals that the bulk spacetime dynamically emerges from the
degrees of freedom on the boundary \cite%
{Rangamani:2016dms,VanRaamsdonk:2010pw,Horowitz:2006ct,Koch:2009gq}. We
further obtain the action of the fluctuation fields on the AdS background at
quadratic and cubic orders.

The organization of the rest of this paper is as follows. In subsection \ref%
{sec_ads_solu}, we analyze the gravity action coupled non-minimally with a
scalar field, with higher-derivative terms, and find AdS solutions. In
subsection \ref{sec_expand}, we expand the fluctuation fields around the AdS
background that we obtained. In subsection \ref{sec_gfgh}, we take care of
the gauge fixing conditions and compute the corresponding ghost action. In
subsections \ref{sec_2nd} and \ref{sec_3rd}, we obtain the action of the
fluctuation fields on the AdS background at quadratic and cubic orders
respectively, which will be relevant to the correlation functions in the
study of AdS/CFT. In subsection \ref{sec_adsreln}, we briefly discuss some
relations to AdS/CFT for these AdS solutions. Finally, we draw some
conclusions in section \ref{sec_conc}.


\section{AdS backgrounds and bulk gravity}

\subsection{Bulk Gravity and AdS}

\label{sec_ads_solu} 

The bulk gravity action is
\begin{equation}
S=\int \mathrm{d}^{d}x\sqrt{-g}\biggl[\frac{1}{2}\xi R\phi ^{2}+aR^{2}-\frac{%
1}{2}(\partial _{\mu }\phi )^{2}-\frac{1}{2}m^{2}\phi ^{2}-V(\phi )\biggr]\,.
\label{eq:bulkact}
\end{equation}%
The total spacetime dimension is $d$. The $\phi $ is the scalar field in the
bulk, $R$ is the Ricci scalar for the corresponding bulk metric $g_{\mu \nu
} $, $m^{2}$ is the mass-squared of the scalar, while the potential $V(\phi
) $ dictates the self-interactions of the scalar. The coupling $\xi $ is
dimensionless and positive in all spacetime dimensions, i.e. $\xi $ $%
\geqslant 0$. The signature taken is $\{-,+,+,+,\cdots \}$, while the
spacetime dimension $d$ is kept arbitrary. The term $\frac{1}{2}\xi R\phi
^{2}~$in the action encodes the explicit non-minimal coupling of the scalar
field $\phi ~$to the Ricci curvature $R$. For the Euclidean path integral to
be bounded from below, we demand that $a\geqslant 0.$

We will study the solutions to the equation of motion of the above action to
look for cases where AdS geometry can be realized. The equation of motion
for the above system can be obtained by varying the action with respect to
fields $g_{\mu \nu }$ and $\phi $. The equation of motion for the action
given in Eq. (\ref{eq:bulkact}) is given by
\begin{eqnarray}
&&a\biggl[\frac{1}{2}g_{\mu \nu }R^{2}-2RR_{\mu \nu }+2\nabla _{\mu }\nabla
_{\nu }R-2g_{\mu \nu }\Box R\biggr]+\frac{1}{2}\xi \phi ^{2}\left( \frac{1}{2%
}g_{\mu \nu }R-R_{\mu \nu }\right)  \notag  \label{eq:eqmFull2} \\
&&+\frac{1}{2}\xi (\nabla _{\mu }\nabla _{\nu }-g_{\mu \nu }\Box )\phi ^{2}-%
\frac{1}{2}\biggl(\frac{1}{2}g_{\mu \nu }(\partial \phi )^{2}-\partial _{\mu
}\phi \partial _{\nu }\phi \biggr)-\frac{1}{2}g_{\mu \nu }\left( \frac{1}{2}%
m^{2}\phi ^{2}+V(\phi )\right) =0\,,
\end{eqnarray}%
and%
\begin{equation}
\Box \phi +\xi R\phi -m^{2}\phi -V^{\prime }(\phi )=0\,.  \label{eq:phi}
\end{equation}%
The first and second equations come by varying the action with respect to
fields $g_{\mu \nu }$ and $\phi $ respectively.

To search for maximally symmetric solutions such as the AdS geometry we take
\begin{equation}
g_{\mu \nu }=\bar{g}_{\mu \nu },~~~R={\bar{R}},~~~~\phi ={\bar{\phi}},
\end{equation}%
where a bar on quantities denotes their background values on the solution.
For these geometries
\begin{equation}
{\bar{R}}_{{\mu \nu }\rho \sigma }=\frac{{\bar{R}}}{d(d-1)}(\bar{g}_{\mu
\rho }\bar{g}_{\nu \sigma }-\bar{g}_{\mu \sigma }\bar{g}_{\nu \rho })\,,%
\hspace{5mm}{\bar{R}}_{\mu \nu }=\frac{{\bar{R}}}{d}\bar{g}_{\mu \nu }\,,
\label{eq:Reim}
\end{equation}%
and ${\bar{R}}$ doesn't depend on spacetime. On AdS spaces ${\bar{R}}<0$,
while ${\bar{R}}>0$ for dS, and ${\bar{R}}=0$ for flat spaces. The equation
of motion simplifies very much if one chooses to look for maximally
symmetric solutions. This gives the background geometry.

For maximally symmetric solutions, the simplified equation of motion for
these is given by,
\begin{eqnarray}
\frac{a(d-4)}{d}{\bar{R}}^{2}+\frac{\xi (d-2)}{2d}{\bar{\phi}}^{2}{\bar{R}}%
-V({\bar{\phi}})-\frac{1}{2}m^{2}{\bar{\phi}}^{2} &=&0\,,
\label{eq:maxsymEQM} \\
\xi {\bar{R}\bar{\phi}}-m^{2}{\bar{\phi}}-V^{\prime }({\bar{\phi}}) &=&0\,.
\end{eqnarray}%
Then we get
\begin{equation}
\bar{R}=\left( m^{2}+\frac{V^{\prime }(\bar{\phi})}{\bar{\phi}}\right) \frac{%
1}{\xi }\,.  \label{bar_R_1}
\end{equation}%
The value $\bar{\phi}$ is the solution to the equation
\begin{equation}
\frac{a(d-4)}{\xi ^{2}d}\left( m^{2}+\frac{V^{\prime }(\bar{\phi})}{\bar{\phi%
}}\right) ^{2}-\frac{1}{d}m^{2}\bar{\phi}^{2}-V(\bar{\phi})+\frac{d-2}{2d}%
V^{\prime }(\bar{\phi})\bar{\phi}=0\,.  \label{eq_bar_phi_01}
\end{equation}%
For AdS solution,%
\begin{equation}
m^{2}+V^{\prime }(\bar{\phi})/\bar{\phi}<0,
\end{equation}%
hence ${\bar{R}}<0~$for the AdS geometry.

The gravitational dynamics is dictated by the non-minimal piece ($R\phi ^{2}$%
). In the case when $\phi $ acquires a nonzero vacuum expectation value
(vev), the action acquires a simple form whose resemblance is that of
Einstein-Hilbert gravity with cosmological constant term. The effective
Newton's constant $G_{N}$ is extracted from the coefficient of the kinetic
term of the metric fluctuation
\begin{equation}
\left( \frac{1}{2}a{\bar{R}}+\frac{1}{8}\xi {\bar{\phi}}^{2}\right) \sqrt{-%
\bar{g}}h_{\mu \nu }\Box h^{\mu \nu }=\frac{1}{16\pi G_{N}}\sqrt{-\bar{g}}~%
\frac{1}{4}h_{\mu \nu }\Box h^{\mu \nu }.
\end{equation}%
(See Eq. (\ref{eq:2nd_order}) for detailed derivation.) The part of the
residual action including the Einstein-Hilbert and the cosmological constant
pieces, can be written as,
\begin{equation}
S=\frac{1}{16\pi G_{N}}\int \mathrm{d}^{d}x\sqrt{-g}\left[ R-\Lambda \right]
\,.  \label{eq:residuAct}
\end{equation}%
So from here one has $\Lambda <0$ when one has an AdS geometry. The Newton's
constant is obtained after the scalar field receives the vev. The
corresponding Newton's constant is
\begin{equation}
G_{N}=\frac{1}{8\pi (4a{\bar{R}}+\xi {\bar{\phi}}^{2})}~.
\end{equation}%
The positivity of the Newton's constant requires that
\begin{equation}
\frac{{\bar{\phi}}^{2}}{|{\bar{R}|}}\geqslant \frac{4a}{\xi }.
\end{equation}%
We have that $G_{N}=l_{p}^{d-2},$ and the mass dimension is $%
[G_{N}^{-1}]=[M^{d-2}]$. The $d$-dimensional Planck constant is
\begin{equation}
l_{p}=\left[ 8\pi (4a{\bar{R}}+\xi {\bar{\phi}}^{2})\right] ^{-1/(d-2)}.
\end{equation}

The radius of curvature of the AdS background is\vspace{1pt}
\begin{equation}
R_{AdS}=\sqrt{\frac{-\xi d(d-1)}{m^{2}+V^{\prime }(\bar{\phi})/\bar{\phi}}}.
\label{eq:RAdS}
\end{equation}%
There is implicit dependence of $R_{AdS}$ on $a$. In the Planck unit,
\begin{equation}
\frac{R_{AdS}}{l_{p}}=\sqrt{\frac{-\xi d(d-1)}{m^{2}+V^{\prime }(\bar{\phi})/%
\bar{\phi}}\left[ 8\pi (4a{\bar{R}}+\xi {\bar{\phi}}^{2})\right] ^{\frac{2}{%
d-2}}}.  \label{eq:RAdSonLp}
\end{equation}%
Our solutions for the metric of the AdS background in Poincare coordinates
is
\begin{equation}
ds^{2}=\frac{R_{AdS}^{2}}{z^{2}}[dz^{2}+\eta _{ab}dx^{a}dx^{b}],
\label{metric}
\end{equation}%
where $z$ and $x^{a}$ with $a=0,1,...,d-2~$are $d$ dimensional coordiantes, $%
\eta _{ab}$ is a metric of $d-1$ dimensional flat spacetime with $%
a,b=0,1,...,d-2$, the radius\ $R_{AdS}$ is (\ref{eq:RAdSonLp}), and ${\bar{R}%
}$ and ${\bar{\phi}}$ are (\ref{bar_R_1}).

We then consider examples of the potential $V(\phi )$ in various dimensions.
In general dimensions including $d=4$, we consider an example
\begin{equation}
V(\phi )=\frac{\lambda }{4}\phi ^{4}+\frac{\mu }{3}\phi ^{3},~~~~~~\lambda
>0.  \label{eq:phi4_phi3}
\end{equation}%
Then using equation (\ref{eq_bar_phi_01}), we get, for $d=4,$%
\begin{eqnarray}
{\bar{\phi}} &=&-\frac{3m^{2}}{\mu }, \\
{\bar{R}} &=&\frac{m^{2}}{\xi }(\frac{9\lambda m^{2}}{\mu ^{2}}-2).
\end{eqnarray}%
For the above AdS solutions with ${\bar{R}<0}${,} the condition on the
parameters is $0<m^{2}<$ $\frac{2\mu ^{2}}{9\lambda }$ for $d=4$.~While for $%
d\neq 4$, the solutions are more complicated looking, and for $d\neq 4,$ $%
a=0,$ we have a simple expression,
\begin{eqnarray}
{\bar{\phi}} &=&-\frac{\mu }{3(d-4)\lambda }\left( d-6+\sqrt{%
(d-6)^{2}+36(d-4)\frac{\lambda m^{2}}{\mu ^{2}}}\right) , \\
{\bar{R}} &=&\frac{d}{\xi (d-4)}(m^{2}+\frac{1}{3}\mu {\bar{\phi})}.
\end{eqnarray}%
For the above AdS solutions with ${\bar{R}<0}${,} the condition on the
parameters is: $0<m^{2}<$ $\frac{2\mu ^{2}}{9\lambda }$ for $4<d\leqslant 6$%
, while $-\frac{(d-6)^{2}\mu ^{2}}{36(d-4)\lambda }\leqslant m^{2}<\frac{%
2\mu ^{2}}{9\lambda }$ for $d>6$. These AdS solutions are obtained when the
scalar field acquires a vacuum expectation value. Here we have shown
explicitly the examples with the potential (\ref{eq:phi4_phi3}). One could
also have more complicated form of $V(\phi ).$

The variation of fields on the background are $\phi =\bar{\phi}+\epsilon
\varphi $ and $g_{\mu \nu }=\bar{g}_{\mu \nu }+\epsilon h_{\mu \nu }$. The
potential can be expanded near $\bar{\phi}$ as
\begin{equation}
{V(\phi )=V({\bar{\phi}})+\epsilon V^{\prime }({\bar{\phi}})\varphi +}\frac{1%
}{2}{\epsilon }^{2}{V}^{\prime \prime }{({\bar{\phi}})\varphi }^{2}+\frac{1}{%
6}{\epsilon }^{3}{V}^{\prime \prime \prime }{({\bar{\phi}})\varphi }^{3}+%
\frac{1}{24}{\epsilon }^{4}{V}^{\prime \prime \prime \prime }{({\bar{\phi}}%
)\varphi }^{4}+\cdots ~.
\end{equation}%
The derivatives of $V(\phi )$ near $\bar{\phi}~$will be used in the next
subsections.

We also mention that more general solutions with non-constant $\bar{\phi}$
from Eq. (\ref{eq:phi}) are more complicated, and one may use the useful
techniques developed in \cite{Schoen:1984,Lee:1987}.


\subsection{Expansions}

\label{sec_expand} 

In the last subsection we have found solutions that has an AdS geometry. For
this we first compute the equation of motion by doing a generic variation of
action with respect to fields. This will result in equation of motion.
Moreover, if we expand the action Eq. (\ref{eq:bulkact}) around the
background dictated by the last subsection, in powers of fluctuation fields,
then it is seen that the linear order terms in the fluctuations will
correspond to equation of motion, the second order terms will give the
kinetic behavior (propagator, masses), while the higher order terms will
give the interactions. Here we will be only concerned with computation up to
third order expansion of action given in Eq. (\ref{eq:bulkact}). In the
following we will give the expansions of various geometrical quantities,
followed by second and third order expansions of the action. These
expansions will be relevant for the two-point and three-point correlators in
AdS/CFT.

Here we give details of the various series expansions that are used in this
paper. The fields are decomposed around a fixed background, where the
background carries the spacetime dependence which is the outcome of equation
of motions of the theory. We have that
\begin{equation}
g_{\mu \nu }=\bar{g}_{\mu \nu }+\epsilon h_{\mu \nu },\,\hspace{5mm}\phi =%
\bar{\phi}+\epsilon \varphi \,.  \label{eq:break}
\end{equation}%
Here $\epsilon $ is a small parameter which is an expansion parameter. It is
incorporated in order to keep track of the expansion-order. Under this
decomposition, the expansion of the inverse metric and the determinant of
metric is given by,
\begin{eqnarray}
&&g^{\mu \nu }=\bar{g}^{\mu \nu }-\epsilon h^{\mu \nu }+\epsilon ^{2}h^{\mu
}{}_{\alpha }h^{\alpha \nu }-\epsilon ^{3}h^{\mu }{}_{\alpha }h^{\alpha
}{}_{\beta }h^{\beta \nu }\cdots ,  \notag  \label{eq:exp_invdet} \\
&&\sqrt{-g}=\sqrt{-\bar{g}}\biggl[1+\frac{\epsilon }{2}h+\epsilon ^{2}\left(
\frac{h^{2}}{8}-\frac{1}{4}h_{\mu \nu }h^{\mu \nu }\right) +\epsilon
^{3}\left( \frac{h^{3}}{48}-\frac{h}{8}h_{\mu \nu }h^{\mu \nu }+\frac{1}{6}%
h^{\mu }{}_{\alpha }h^{\alpha }{}_{\beta }h^{\beta }{}_{\mu }\right) +\cdots %
\biggr]\,,  \notag \\
&&
\end{eqnarray}%
where $h=\bar{g}^{\mu \nu }h_{\mu \nu }$. The expansion of Christoffel
connection can be obtained by plugging the expansion of metric and its
inverse in the definition of the Christoffel connection, and collecting
pieces at various orders of $\epsilon $. This gives
\begin{eqnarray}
\Gamma _{\alpha }{}^{\mu }{}_{\beta } &=&\bar{\Gamma}_{\alpha }{}^{\mu
}{}_{\beta }+\epsilon \left( \bar{\nabla}_{\alpha }h^{\mu }{}_{\beta }+\bar{%
\nabla}_{\beta }h^{\mu }{}_{\alpha }-\bar{\nabla}^{\mu }h_{\alpha \beta
}\right) -\epsilon ^{2}h^{\mu \rho }\left( \bar{\nabla}_{\alpha }h_{\rho
\beta }+\bar{\nabla}_{\beta }h_{\rho \alpha }-\bar{\nabla}_{\rho }h_{\alpha
\beta }\right)  \notag  \label{eq:Conexp} \\
&&+\epsilon ^{3}h^{\mu }{}_{\sigma }h^{\sigma \rho }\left( \bar{\nabla}%
_{\alpha }h_{\rho \beta }+\bar{\nabla}_{\beta }h_{\rho \alpha }-\bar{\nabla}%
_{\rho }h_{\alpha \beta }\right) +\cdots \,.
\end{eqnarray}%
From here the variation of Christoffel connection at various orders can be
read easily. Using this one can perform a series expansion of the various
curvature tensors: Riemann tensor, Ricci tensor and Ricci scalar. We define
the Riemann tensor in the following manner,
\begin{equation}
\lbrack \nabla _{\mu },\nabla _{\nu }]V^{\rho }=R_{\mu \nu }{}^{\rho
}{}_{\sigma }V^{\sigma }\,,\hspace{5mm}R_{\mu \nu }{}^{\rho }{}_{\sigma
}=\partial _{\mu }\Gamma _{\nu }{}^{\rho }{}_{\sigma }-\partial _{\nu
}\Gamma _{\mu }{}^{\rho }{}_{\sigma }+\Gamma _{\mu }{}^{\rho }{}_{\lambda
}\Gamma _{\nu }{}^{\lambda }{}_{\sigma }-\Gamma _{\nu }{}^{\rho }{}_{\lambda
}\Gamma _{\mu }{}^{\lambda }{}_{\sigma }\,,  \label{eq:Riemanndef}
\end{equation}%
where $V^{\mu }$ is an arbitrary vector field. The expansion of the Riemann
tensor is given by,
\begin{eqnarray}
&&R_{\mu \nu }{}^{\rho }{}_{\sigma }=\bar{R}_{\mu \nu }{}^{\rho }{}_{\sigma
}+\epsilon \left( \bar{\nabla}_{\mu }\Gamma _{\nu }^{(1)}{}^{\rho
}{}_{\sigma }-\bar{\nabla}_{\nu }\Gamma _{\mu }^{(1)}{}^{\rho }{}_{\sigma
}\right)  \notag  \label{eq:Riemannexp} \\
&&+\epsilon ^{2}\biggl(\bar{\nabla}_{\mu }\Gamma _{\nu }^{(2)}{}^{\rho
}{}_{\sigma }-\bar{\nabla}_{\nu }\Gamma _{\mu }^{(2)}{}^{\rho }{}_{\sigma
}+\Gamma _{\mu }^{(1)}{}^{\rho }{}_{\lambda }\Gamma _{\nu }^{(1)}{}^{\lambda
}{}_{\sigma }-\Gamma _{\nu }^{(1)}{}^{\rho }{}_{\lambda }\Gamma _{\mu
}^{(1)}{}^{\lambda }{}_{\sigma }\biggr)  \notag \\
&&+\epsilon ^{3}\biggl(\bar{\nabla}_{\mu }\Gamma _{\nu }^{(3)}{}^{\rho
}{}_{\sigma }-\bar{\nabla}_{\nu }\Gamma _{\mu }^{(3)}{}^{\rho }{}_{\sigma
}+\Gamma _{\mu }^{(1)}{}^{\rho }{}_{\lambda }\Gamma _{\nu }^{(2)}{}^{\lambda
}{}_{\sigma }+\Gamma _{\mu }^{(2)}{}^{\rho }{}_{\lambda }\Gamma _{\nu
}^{(1)}{}^{\lambda }{}_{\sigma }-\Gamma _{\nu }^{(1)}{}^{\rho }{}_{\lambda
}\Gamma _{\mu }^{(2)}{}^{\lambda }{}_{\sigma }-\Gamma _{\nu }^{(2)}{}^{\rho
}{}_{\lambda }\Gamma _{\mu }^{(1)}{}^{\lambda }{}_{\sigma }\biggr)+\cdots \,.
\notag \\
&&
\end{eqnarray}%
The Ricci tensor is then obtained by contracting the first and third index
of the above expression of the Riemann tensor. Its expansion is given by,
\begin{eqnarray}
&&R_{\nu \sigma }=R_{\rho \nu }{}^{\rho }{}_{\sigma }=\bar{R}_{\nu \sigma
}+\epsilon \left( \bar{\nabla}_{\rho }\Gamma _{\nu }^{(1)}{}^{\rho
}{}_{\sigma }-\bar{\nabla}_{\nu }\Gamma _{\rho }^{(1)}{}^{\rho }{}_{\sigma
}\right)  \notag  \label{eq:RicciTexp} \\
&&+\epsilon ^{2}\biggl(\bar{\nabla}_{\rho }\Gamma _{\nu }^{(2)}{}^{\rho
}{}_{\sigma }-\bar{\nabla}_{\nu }\Gamma _{\rho }^{(2)}{}^{\rho }{}_{\sigma
}+\Gamma _{\rho }^{(1)}{}^{\rho }{}_{\lambda }\Gamma _{\nu
}^{(1)}{}^{\lambda }{}_{\sigma }-\Gamma _{\nu }^{(1)}{}^{\rho }{}_{\lambda
}\Gamma _{\rho }^{(1)}{}^{\lambda }{}_{\sigma }\biggr)  \notag \\
&&+\epsilon ^{3}\biggl(\bar{\nabla}_{\rho }\Gamma _{\nu }^{(3)}{}^{\rho
}{}_{\sigma }-\bar{\nabla}_{\nu }\Gamma _{\rho }^{(3)}{}^{\rho }{}_{\sigma
}+\Gamma _{\rho }^{(1)}{}^{\rho }{}_{\lambda }\Gamma _{\nu
}^{(2)}{}^{\lambda }{}_{\sigma }+\Gamma _{\rho }^{(2)}{}^{\rho }{}_{\lambda
}\Gamma _{\nu }^{(1)}{}^{\lambda }{}_{\sigma }-\Gamma _{\nu }^{(1)}{}^{\rho
}{}_{\lambda }\Gamma _{\rho }^{(2)}{}^{\lambda }{}_{\sigma }-\Gamma _{\nu
}^{(2)}{}^{\rho }{}_{\lambda }\Gamma _{\rho }^{(1)}{}^{\lambda }{}_{\sigma }%
\biggr)+\cdots \,.  \notag \\
&&
\end{eqnarray}%
From this we note that a useful handy expression that often enters these
expansion. These expression in terms of fluctuation $h_{\mu \nu }$ can be
written in simple forms as,
\begin{equation}
\Gamma _{\rho }^{(1)}{}^{\rho }{}_{\sigma }=\frac{1}{2}\bar{\nabla}_{\sigma
}h\,,\hspace{5mm}\Gamma _{\rho }^{(2)}{}^{\rho }{}_{\sigma }=-\frac{1}{2}%
h^{\rho \alpha }\bar{\nabla}_{\sigma }h_{\rho \alpha }\,,\hspace{5mm}\Gamma
_{\rho }^{(3)}{}^{\rho }{}_{\sigma }=\frac{1}{2}h^{\rho }{}_{\theta
}h^{\theta \alpha }\bar{\nabla}_{\sigma }h_{\rho \alpha }~.
\label{eq:GammaExp}
\end{equation}%
One can use these to obtain the expansion of the Ricci tensor. At each order
one can perform commutation of covariant derivatives to obtain expressions
which involve divergences of $h_{\mu \nu }$. At first order it is given by,
\begin{equation}
R_{\nu \sigma }^{(1)}=\frac{1}{2}\biggl[\bar{\nabla}_{\nu }\bar{\nabla}%
_{\rho }h^{\rho }{}_{\sigma }+\bar{\nabla}_{\sigma }\bar{\nabla}_{\rho
}h^{\rho }{}_{\nu }+\bar{R}_{\nu \lambda }h^{\lambda }{}_{\sigma }+\bar{R}%
_{\sigma \lambda }h^{\lambda }{}_{\nu }-2\bar{R}_{\rho \nu }{}^{\lambda
}{}_{\sigma }h^{\rho }{}_{\lambda }-\Box h_{\nu \sigma }-\bar{\nabla}_{\nu }%
\bar{\nabla}_{\sigma }h\biggr].
\end{equation}%
At second order the expansion is given by,
\begin{eqnarray}
&&R_{\mu \nu }^{(2)}=\bar{\nabla}_{\rho }\Gamma _{\mu }^{(2)}{}^{\rho
}{}_{\nu }-\bar{\nabla}_{\mu }\Gamma _{\rho }^{(2)}{}^{\rho }{}_{\nu }+\frac{%
1}{4}\biggl[\bar{\nabla}^{\lambda }h\left( \bar{\nabla}_{\mu }h_{\lambda \nu
}+\bar{\nabla}_{\nu }h_{\lambda \mu }-\bar{\nabla}_{\lambda }h_{\mu \nu
}\right) -\bar{\nabla}_{\mu }h^{\rho \lambda }\bar{\nabla}_{\nu }h_{\rho
\lambda }  \notag  \label{eq:RicciT2} \\
&&-\bar{\nabla}^{\lambda }h^{\rho }{}_{\mu }\bar{\nabla}_{\rho }h_{\lambda
\nu }+\bar{\nabla}^{\lambda }h^{\rho }{}_{\mu }\bar{\nabla}_{\lambda
}h_{\rho \nu }+\bar{\nabla}^{\rho }h_{\mu }{}^{\lambda }\bar{\nabla}_{\rho
}h_{\lambda \nu }-\bar{\nabla}^{\rho }h_{\mu }{}^{\lambda }\bar{\nabla}%
_{\lambda }h_{\rho \nu }\biggr]\,.
\end{eqnarray}%
At third order the expansion is given by,
\begin{eqnarray}
&&R_{\mu \nu }^{(3)}=\bar{\nabla}_{\rho }\Gamma _{\mu }^{(3)}{}^{\rho
}{}_{\nu }-\bar{\nabla}_{\mu }\Gamma _{\rho }^{(3)}{}^{\rho }{}_{\nu }-\frac{%
1}{4}(\bar{\nabla}_{\alpha }h)h^{\alpha \lambda }\left( \bar{\nabla}_{\mu
}h_{\lambda \nu }+\bar{\nabla}_{\nu }h_{\lambda \mu }-\bar{\nabla}_{\lambda
}h_{\mu \nu }\right)  \notag  \label{eq:RicciT3} \\
&&-\frac{1}{4}(h^{\rho \alpha }\bar{\nabla}^{\lambda }h_{\rho \alpha
})\left( \bar{\nabla}_{\mu }h_{\lambda \nu }+\bar{\nabla}_{\nu }h_{\lambda
\mu }-\bar{\nabla}_{\lambda }h_{\mu \nu }\right) -\frac{1}{2}h^{\lambda
\alpha }\biggl[\bar{\nabla}_{\mu }h^{\rho }{}_{\lambda }\bar{\nabla}_{\nu
}h_{\alpha \rho }+\bar{\nabla}_{\lambda }h^{\rho }{}_{\mu }\bar{\nabla}_{\nu
}h_{\alpha \rho }  \notag \\
&&-\bar{\nabla}^{\rho }h_{\mu \lambda }\bar{\nabla}_{\rho }h_{\alpha \nu }+%
\bar{\nabla}^{\rho }h_{\mu \lambda }\bar{\nabla}_{\alpha }h_{\rho \nu }%
\biggr]\,.
\end{eqnarray}%
In order to obtain the expansion of the Ricci scalar one has to make use of
both the expansion of the inverse metric and Ricci tensor. At first order
this expansion is given by,
\begin{equation}
R^{(1)}=\bar{\nabla}_{\mu }\bar{\nabla}_{\nu }h^{\mu \nu }-\bar{\Box}h-\bar{R%
}_{\mu \nu }h^{\mu \nu }\,.  \label{eq:Ricci1}
\end{equation}%
At second order we have the following,
\begin{eqnarray}
&&R^{(2)}=\bar{R}_{\rho \nu \lambda \sigma }h^{\rho \lambda }h^{\nu \sigma
}-h^{\nu \sigma }\bar{\nabla}_{\nu }\bar{\nabla}_{\rho }h^{\rho }{}_{\sigma
}+\frac{1}{2}h^{\nu \sigma }\bar{\Box}h_{\nu \sigma }+\frac{1}{2}h^{\nu
\sigma }\bar{\nabla}_{\nu }\bar{\nabla}_{\sigma }h+\frac{1}{2}\bar{\nabla}%
^{\lambda }h\bar{\nabla}^{\sigma }h_{\lambda \sigma }  \notag
\label{eq:Ricci2} \\
&&-\frac{1}{4}\bar{\nabla}^{\lambda }h\bar{\nabla}_{\lambda }h+\frac{1}{4}%
\bar{\nabla}^{\rho }h^{\nu \lambda }\bar{\nabla}_{\rho }h_{\nu \lambda }-%
\frac{1}{2}\bar{\nabla}^{\lambda }h^{\rho \nu }\bar{\nabla}_{\rho
}h_{\lambda \nu }+\bar{g}^{\nu \sigma }\left( \bar{\nabla}_{\rho }\Gamma
_{\nu }^{(2)}{}^{\rho }{}_{\sigma }-\bar{\nabla}_{\nu }\Gamma _{\rho
}^{(2)}{}^{\rho }{}_{\sigma }\right) \,.
\end{eqnarray}%
At third order the expansion is more elaborate as there are more number of
terms. We have,
\begin{eqnarray}
&&R^{(3)}=\bar{g}^{\nu \sigma }\left( \bar{\nabla}_{\rho }\Gamma _{\nu
}^{(3)}{}^{\rho }{}_{\sigma }-\bar{\nabla}_{\nu }\Gamma _{\rho
}^{(3)}{}^{\rho }{}_{\sigma }\right) -\frac{1}{4}h^{\lambda \alpha }(\bar{%
\nabla}_{\lambda }h)\left( 2\bar{\nabla}^{\nu }h_{\nu \alpha }-\bar{\nabla}%
_{\alpha }h\right)  \notag  \label{eq:Ricci3} \\
&&-\frac{1}{4}(h_{\rho \lambda }\bar{\nabla}^{\alpha }h^{\rho \lambda
})\left( 2\bar{\nabla}^{\nu }h_{\nu \alpha }-\bar{\nabla}_{\alpha }h\right) -%
\frac{1}{2}h^{\lambda \alpha }\biggl(\bar{\nabla}_{\nu }h^{\rho }{}_{\lambda
}\bar{\nabla}^{\nu }h_{\alpha \rho }+\bar{\nabla}_{\lambda }h^{\rho }{}_{\nu
}\bar{\nabla}^{\nu }h_{\alpha \rho }-\bar{\nabla}^{\rho }h_{\nu \lambda }%
\bar{\nabla}_{\rho }h_{\alpha }{}^{\nu }  \notag \\
&&+\bar{\nabla}^{\rho }h_{\nu \lambda }\bar{\nabla}_{\alpha }h_{\rho
}{}^{\nu }\biggr)-h^{\nu }{}_{\theta }h^{\theta }{}_{\alpha }h^{\alpha
\sigma }\bar{R}_{\nu \sigma }+\frac{1}{2}h^{\nu }{}_{\alpha }h^{\alpha
\sigma }\biggl(\bar{\nabla}_{\nu }\bar{\nabla}_{\rho }h^{\rho }{}_{\sigma }+%
\bar{\nabla}_{\sigma }\bar{\nabla}_{\rho }h^{\rho }{}_{\nu }+\bar{R}_{\nu
\lambda }h^{\lambda }{}_{\sigma }+\bar{R}_{\sigma \lambda }h^{\lambda
}{}_{\nu }  \notag \\
&&-2\bar{R}_{\rho \nu }{}^{\lambda }{}_{\sigma }h^{\rho }{}_{\lambda }-\Box
h_{\nu \sigma }-\bar{\nabla}_{\nu }\bar{\nabla}_{\sigma }h\biggr)-h^{\mu \nu
}(\bar{\nabla}_{\rho }\Gamma _{\mu }^{(2)}{}^{\rho }{}_{\nu }-\bar{\nabla}%
_{\mu }\Gamma _{\rho }^{(2)}{}^{\rho }{}_{\nu })+\frac{1}{4}h^{\mu \nu }%
\biggl[\bar{\nabla}^{\lambda }h\bigl(\bar{\nabla}_{\mu }h_{\lambda \nu }+%
\bar{\nabla}_{\nu }h_{\lambda \mu }  \notag \\
&&-\bar{\nabla}_{\lambda }h_{\mu \nu }\bigr)-\bar{\nabla}_{\mu }h^{\rho
\lambda }\bar{\nabla}_{\nu }h_{\rho \lambda }-\bar{\nabla}^{\lambda }h^{\rho
}{}_{\mu }\bar{\nabla}_{\rho }h_{\lambda \nu }+\bar{\nabla}^{\lambda
}h^{\rho }{}_{\mu }\bar{\nabla}_{\lambda }h_{\rho \nu }+\bar{\nabla}^{\rho
}h_{\mu }{}^{\lambda }\bar{\nabla}_{\rho }h_{\lambda \nu }-\bar{\nabla}%
^{\rho }h_{\mu }{}^{\lambda }\bar{\nabla}_{\lambda }h_{\rho \nu }\biggr]\,.
\notag \\
&&
\end{eqnarray}


\subsection{Gauge-fixing and Faddeev-Popov ghosts}

\label{sec_gfgh} 

Here in this subsection we will take into account the gauge fixing condition
and compute the corresponding ghost action.

We analyze the diffeomorphism invariant action of the coupled system using
background field method \cite{DeWitt3, Abbott}. It is advantageous, as by
construction it preserves background gauge invariance. The field is
decomposed into background and fluctuation. Keeping the background fixed the
path-integral is then reduced to an integral over the fluctuations. The
gravitational metric field is decomposed into background and fluctuation. To
prevent over-counting of gauge-orbits in the path-integral measure, a
constraint is applied on this fluctuation field, which results in appearance
of auxiliary fields called ghosts. This procedure of systematically applying
the constraint leading to ghost can be elegantly taken care of by the
Faddeev-Popov prescription \cite{Faddeev:1967fc,DeWitt:1964yg}. The
effective action generated after integrating over the fluctuation and
auxiliary fields still enjoys invariance over the background fields.

The diffeomorphism invariance of the full action in Eq. (\ref{eq:bulkact})
implies that for arbitrary vector field $V^{\rho }$, the action should be
invariant under the following transformation of the metric field variable,
\begin{equation}
\delta _{D}\gamma _{\mu \nu }=\mathcal{L}_{V}\gamma _{\mu \nu }=V^{\rho
}\partial _{\rho }\gamma _{\mu \nu }+\gamma _{\mu \rho }\partial _{\nu
}V^{\rho }+\gamma _{\nu \rho }\partial _{\mu }V^{\rho }\,,
\label{eq:gaugetrgamma}
\end{equation}%
where $\mathcal{L}_{V}\gamma _{\mu \nu }$ is the Lie derivative of the
quantum metric $\gamma _{\mu \nu }$ along the vector field $V^{\rho }$.
Decomposing the quantum metric $\gamma _{\mu \nu }$ into background ($\bar{g}%
_{\mu \nu }$) and fluctuation ($\epsilon h_{\mu \nu }$) allows one to figure
out the transformation of the fluctuation field while keeping the background
fixed. This will imply the following transformation of $h_{\mu \nu }$,
\begin{equation}
\delta _{D}h_{\mu \nu }=\frac{1}{\epsilon }\left( \bar{\nabla}_{\mu }V_{\nu
}+\bar{\nabla}_{\nu }V_{\mu }\right) +V^{\rho }\bar{\nabla}_{\rho }h_{\mu
\nu }+h_{\mu \rho }\bar{\nabla}_{\nu }V^{\rho }+h_{\nu \rho }\bar{\nabla}%
_{\mu }V^{\rho }\,,  \label{eq:gaugetrh}
\end{equation}%
where $\bar{\nabla}$ is the covariant derivative whose connection is
constructed using the background metric. This is the full transformation of
the metric fluctuation field. For small $\epsilon $ only the leading term is
relevant. In the quantum theory it is seen that the leading term leads to
one-loop effects while higher-loop comes from $\mathcal{O}(\epsilon ^{0})$
terms. The invariance of the action is broken by choosing an appropriate
gauge-fixing condition implemented via Faddeev-Popov procedure \cite%
{Faddeev:1967fc}.

The gauge fixing action chosen for fixing the invariance under the
transformation of the metric fluctuation field is given by,
\begin{equation}
S_{\mathrm{gf}}=\frac{1}{2\alpha }\int \,\mathrm{d}^{d}x\sqrt{-\bar{g}}(\bar{%
\nabla}^{\rho }h_{\rho \mu })\bar{g}^{\mu \nu }(\bar{\nabla}^{\sigma
}h_{\sigma \nu })\,.  \label{eq:Sgfact}
\end{equation}%
This gauge-fixing action introduces the gauge-fixing parameter $\alpha $.
When $\alpha =0$ the gauge condition is imposed sharply, resulting in Landau
gauge $F_{\nu }=\bar{\nabla}^{\mu }h_{\mu \nu }=0$. The breaking of
gauge-invariance leads to Faddeev-Popov ghosts which is obtained following
the prescription given in \cite{Faddeev:1967fc}. We introduce
gauge-condition in the path-integral by multiplying the later with unity in
the following form,
\begin{equation}
1=\int \mathcal{D}F_{\mu }^{V}\exp \biggl[\frac{i}{2\alpha }\int \mathrm{d}%
^{d}x\sqrt{-\bar{g}}F_{\mu }^{V}\bar{g}^{\mu \nu }F_{\nu }^{V}\biggr]\,,
\label{eq:pathunity}
\end{equation}%
where $F_{\mu }^{V}$ is the gauge transformed $F_{\mu }$. The original
path-integral (without gauge-fixing) is invariant under transformation Eq. (%
\ref{eq:gaugetrh}) of the field $h_{\mu \nu }$, and this implies that a
change of integration variable from $h_{\mu \nu }$ to $h_{\mu \nu }^{V}$
doesn't give rise to any Jacobian in the path-integral measure. However
replacing the measure over $F_{\mu }^{V}$ with the measure over $V^{\rho }$
introduces a non-trivial Jacobian in the path-integral. This is obtained as
follows,
\begin{equation}
\mathcal{D}F_{\mu }^{V}=\det \biggl(\frac{\partial F_{\mu }}{\partial
V^{\rho }}\biggr)\mathcal{D}V^{\rho }\,.  \label{eq:gaugejacob1}
\end{equation}%
In the background field formalism this Jacobian consists of background
covariant derivative, background and fluctuation fields, and is independent
of the transformation parameter $V^{\rho }$. This implies that it can be
taken out of the functional integral over $V^{\rho }$. Changing the
integration variable from $h_{\mu \nu }^{V}$ to $h_{\mu \nu }$, and ignoring
the infinite constant generated by integrating over $V^{\rho }$, gives us
the gauge fixed path integral.

The Faddeev-Popov determinant in Eq. (\ref{eq:gaugejacob1}) is then
exponentiated by making use of anti-commuting auxiliary fields. These
auxiliary fields are known as Faddeev-Popov ghosts. The path integral of the
full ghost sector is given by,
\begin{equation}
\int \mathcal{D}\bar{C}_{\mu }\mathcal{D}C_{\nu }\,\,\exp \biggl[-i\int
\mathrm{d}^{d}x\sqrt{-\bar{g}}\biggl\{\bar{C}_{\mu }\left( \frac{\partial
F_{\nu }}{\partial V^{\rho}}\right) C^{\rho }\biggr\}\biggr]\,,
\label{eq:ghostpath}
\end{equation}%
where $\bar{C}_{\mu }$ and $C_{\nu }$ are Faddeev-Popov ghost fields arising
from the gauge fixing in the gravitational sector.

In the case when $F_{\mu }$ is given as in Eq. (\ref{eq:Sgfact}), the
Faddeev-Popov ghost action is given by,
\begin{equation}
S_{\mathrm{gh}}^{FP}=-\int \mathrm{d}^{d}x\sqrt{-\bar{g}}\bar{C}_{\mu
}X_{\rho }^{\mu }C^{\rho }\,,  \label{eq:ghostact}
\end{equation}%
where,
\begin{eqnarray}
X_{\rho }^{\mu } &=&\bar{g}^{\mu \nu }\biggl[\frac{1}{\epsilon }\left( \bar{%
\nabla}_{\rho }\bar{\nabla}_{\nu }+\bar{g}_{\nu \rho }\bar{\Box}\right) +%
\bar{\nabla}_{\rho }h_{\sigma \nu }\bar{\nabla}^{\sigma }+\bar{\nabla}%
^{\sigma }\bar{\nabla}_{\rho }h_{\sigma \nu }+\bar{\nabla}^{\sigma }h_{\nu
\rho }\bar{\nabla}_{\sigma }+h_{\nu \rho }\bar{\Box}  \notag
\label{eq:ghostet} \\
&&+\bar{\nabla}^{\sigma }h_{\sigma \rho }\bar{\nabla}_{\nu }+h_{\sigma \rho }%
\bar{\nabla}^{\sigma }\bar{\nabla}_{\nu }\biggr]\,.
\end{eqnarray}%
Here the last several terms contain terms linear in $h_{\mu \nu }$. These
are not relevant in doing one-loop computations, but at higher-loops
they are important. The full action of the theory (which includes the
coupled gravity and matter action, gauge fixing and ghost action) possesses
BRS invariance. Furthermore, in flat spacetime there is a residual symmetry
coming from dilatation invariance \cite{Antoniadis:1985ub}. One can still
make the following coordinate transformation
\begin{equation}
\omega ^{\mu }(x)=\omega x^{\mu }\,,  \label{eq:dilaton}
\end{equation}%
where $\omega $ is a constant parameter. In flat spacetime this is a
dilatation. This symmetry invariance is broken by imposing $h=0$, where $h=%
\mathrm{tr}h_{\nu }^{\mu }$. This doesn't lead to generation of ghost action
\cite{Ohta:2015zwa}. Furthermore, on AdS background with fixed AdS radius,
dilatation invariance is broken. In curved space following \cite%
{Ohta:2015zwa} we use $h=0$ as the trace-free condition. In fact this can be
incorporated in the action of the theory by using a Lagrange multiplier \cite%
{Antoniadis:1985ub}. Variation with respect to Lagrange multiplier leads to
constraint $h=0$. This can be viewed as a constraint imposed to break
dilatation invariance. In the following we will impose the Landau gauge ($%
\bar{\nabla}^{\mu }h_{\mu \nu }=0$) and dilatation breaking ($h=0$) to
obtain the second and third order terms on the maximally symmetric
background.


\subsection{Second order}

\label{sec_2nd} 

In this subsection we compute the second order terms for the action given in
Eq. (\ref{eq:bulkact}). The full second variation on a general background
without the usage of gauge conditions is a complicated expression consisting
of several terms. This acquires a bit simpler form once integration by parts
is done. To compute second order variation of the action one needs the
expansion up to second order in $\epsilon $ of $\sqrt{-g}$, $g^{\mu \nu }$, $%
g_{\mu \nu }$, $\Gamma _{\mu }{}^{\rho }{}_{\nu }$, and $R$. These
expansions are computed in subsection \ref{sec_expand}. Here we make use of
these expansions to compute the variations of the action, on the background
in subsection \ref{sec_ads_solu}. The second variation of the $R\phi ^{2}$
term, without integration by parts, is given by,
\begin{eqnarray}
&&S_{R\phi ^{2}}=\frac{1}{2}\epsilon ^{2}\xi \int \mathrm{d}^{d}x\sqrt{-\bar{%
g}}\biggl[\left( \frac{1}{8}h^{2}-\frac{1}{4}h_{\mu \nu }h^{\mu \nu }\right)
\bar{R}\bar{\phi}^{2}+\bar{\phi}^{2}\biggl\{\bar{R}_{\rho \nu \lambda \sigma
}h^{\rho \lambda }h^{\nu \sigma }-h^{\nu \sigma }\bar{\nabla}_{\nu }\bar{%
\nabla}_{\rho }h^{\rho }{}_{\sigma }  \notag  \label{eq:2varRphi2} \\
&&+\frac{1}{2}h^{\mu \nu }\Box h_{\mu \nu }+\frac{1}{2}h^{\nu \sigma }\bar{%
\nabla}_{\nu }\bar{\nabla}_{\sigma }h+\frac{1}{2}\bar{\nabla}^{\lambda }h%
\bar{\nabla}^{\sigma }h_{\lambda \sigma }-\frac{1}{4}\bar{\nabla}^{\lambda }h%
\bar{\nabla}_{\lambda }h+\frac{1}{4}\bar{\nabla}^{\rho }h^{\nu \lambda }\bar{%
\nabla}_{\rho }h_{\nu \lambda }  \notag \\
&&-\frac{1}{2}\bar{\nabla}^{\lambda }h^{\rho \nu }\bar{\nabla}_{\rho }h_{\nu
\lambda }+\frac{1}{2}h(\bar{\nabla}_{\mu }\bar{\nabla}_{\nu }h^{\mu \nu
}-\Box h-\bar{R}_{\mu \nu }h^{\mu \nu })+\bar{g}^{\nu \sigma }(\bar{\nabla}%
_{\rho }\Gamma _{\nu }^{(2)}{}^{\rho }{}_{\sigma }-\bar{\nabla}_{\nu }\Gamma
_{\rho }^{(2)}{}^{\rho }{}_{\sigma })\biggr\}  \notag \\
&&+\bar{\phi}\biggl\{h\bar{R}\varphi +2\varphi (\bar{\nabla}_{\mu }\bar{%
\nabla}_{\nu }h^{\mu \nu }-\Box h-\bar{R}_{\mu \nu }h^{\mu \nu })\biggr\}+%
\bar{R}\varphi ^{2}\biggr]\,,
\end{eqnarray}%
where the expansions of Christoffel appear in form of total derivative and
won't contribute in the bulk studies. However they generate a surface term.
The second variation of $R^{2}$ term of the action is given by,
\begin{eqnarray}
&&S_{R^{2}}=a\epsilon ^{2}\int \mathrm{d}^{d}x\sqrt{-\bar{g}}\biggl[\left(
\frac{1}{8}h^{2}-\frac{1}{4}h_{\mu \nu }h^{\mu \nu }\right) \bar{R}^{2}+\bar{%
R}h(\bar{\nabla}_{\mu }\bar{\nabla}_{\nu }h^{\mu \nu }-\Box h-\bar{R}_{\mu
\nu }h^{\mu \nu })  \notag  \label{eq:2ndvarR2} \\
&&+\bar{\nabla}_{\mu }\bar{\nabla}_{\nu }h^{\mu \nu }\bar{\nabla}_{\alpha }%
\bar{\nabla}_{\beta }h^{\alpha \beta }+\Box h\Box h+\bar{R}_{\mu \nu }h^{\mu
\nu }\bar{R}_{\alpha \beta }h^{\alpha \beta }-2\bar{\nabla}_{\mu }\bar{\nabla%
}_{\nu }h^{\mu \nu }\Box h-2\bar{\nabla}_{\mu }\bar{\nabla}_{\nu }h^{\mu \nu
}\bar{R}_{\alpha \beta }h^{\alpha \beta }  \notag \\
&&+2\Box h\bar{R}_{\alpha \beta }h^{\alpha \beta }+2\bar{R}\biggl\{\bar{R}%
_{\rho \nu \lambda \sigma }h^{\rho \lambda }h^{\nu \sigma }-h^{\nu \sigma }%
\bar{\nabla}_{\nu }\bar{\nabla}_{\rho }h^{\rho }{}_{\sigma }+\frac{1}{2}%
h^{\mu \nu }\Box h_{\mu \nu }+\frac{1}{2}h^{\nu \sigma }\bar{\nabla}_{\nu }%
\bar{\nabla}_{\sigma }h+\frac{1}{2}\bar{\nabla}^{\lambda }h\bar{\nabla}%
^{\sigma }h_{\lambda \sigma }  \notag \\
&&-\frac{1}{4}\bar{\nabla}^{\lambda }h\bar{\nabla}_{\lambda }h+\frac{1}{4}%
\bar{\nabla}^{\rho }h^{\nu \lambda }\bar{\nabla}_{\rho }h_{\nu \lambda }-%
\frac{1}{2}\bar{\nabla}^{\lambda }h^{\rho \nu }\bar{\nabla}_{\rho }h_{\nu
\lambda }+\frac{1}{2}h(\bar{\nabla}_{\mu }\bar{\nabla}_{\nu }h^{\mu \nu
}-\Box h-\bar{R}_{\mu \nu }h^{\mu \nu })  \notag \\
&&+\bar{g}^{\nu \sigma }(\bar{\nabla}_{\rho }\Gamma _{\nu }^{(2)}{}^{\rho
}{}_{\sigma }-\bar{\nabla}_{\nu }\Gamma _{\rho }^{(2)}{}^{\rho }{}_{\sigma })%
\biggr\}\biggr]\,.
\end{eqnarray}%
The second variation for the matter part is a bit simpler as the
gravitational couplings are simpler. These are given by,
\begin{eqnarray}
&&S_{\mathrm{matter}}=\epsilon ^{2}\int \mathrm{d}^{d}x\sqrt{-\bar{g}}\biggl[%
-\left( \frac{1}{8}h^{2}-\frac{1}{4}h_{\mu \nu }h^{\mu \nu }\right) \left(
\frac{1}{2}m^{2}\bar{\phi}^{2}+V(\bar{\phi})\right) -\frac{1}{2}h(m^{2}\bar{%
\phi}+V^{\prime }(\bar{\phi}))\varphi  \notag  \label{eq:2varMat} \\
&&+\frac{1}{2}\varphi \Box \varphi -\frac{1}{2}(m^{2}+V^{\prime \prime }(%
\bar{\phi}))\varphi ^{2}\biggr]\,.
\end{eqnarray}%
These second order expansion terms are over generic background (with
arbitrary background curvature but with constant $\bar{\phi}$) and in
arbitrary dimensions. However this gets simplified once integration by parts
is done and Landau gauge ($\bar{\nabla}^{\mu }h_{\mu \nu }=0$) and trace
free condition ($h=0$) is employed. Under these constraints the second
variation gets very simplified as many terms go away. Then the residual
second variation is given by,
\begin{eqnarray}
&&S^{(2)}=\epsilon ^{2}\int \mathrm{d}^{d}x\sqrt{-\bar{g}}\biggl[h_{\mu \nu }%
\biggl\{\frac{1}{2}\xi \bar{\phi}^{2}\biggl(\frac{1}{4}\Box -\frac{%
(d^{2}-3d+4)\bar{R}}{4d(d-1)}\biggr)+a\bar{R}\biggl(\frac{1}{2}\Box -\frac{%
(d^{2}-5d+8)\bar{R}}{4d(d-1)}\biggr)  \notag \\
&&+\frac{1}{4}\biggl(\frac{1}{2}m^{2}\bar{\phi}^{2}+V(\bar{\phi})\biggr)%
\biggr\}h^{\mu \nu }+\frac{1}{2}\varphi \left( \Box +\xi \bar{R}%
-m^{2}-V^{\prime \prime }(\bar{\phi})\right) \varphi \biggr]\,.
\label{eq:2nd_order}
\end{eqnarray}%
The surface terms produced during this are included in appendix \ref{bound}.


\subsection{Third order}

\label{sec_3rd} 

The third order terms are needed to compute the cubic couplings. These will
give the three-point correlators on the boundary of AdS spacetime, which
will be relevant for AdS/CFT. At the third order the number of terms are
quite many for each term of the action when expanded. We would make use of
gauge conditions and requirement of maximally symmetric spacetime
background. The third order terms coming from $R\phi ^{2}$ piece of the
action are
\begin{equation}
S_{R\phi ^{2}}^{(3)}=\frac{\epsilon ^{3}}{2}\xi \int \mathrm{d}^{d}x\sqrt{-%
\bar{g}}\biggl[\bar{\phi}^{2}\biggl\{\frac{1}{6}h^{\mu }{}_{\alpha
}h^{\alpha }{}_{\beta }h^{\beta }{}_{\mu }\bar{R}+R^{(3)}\biggr\}+2\bar{\phi}%
\varphi \left( -\frac{1}{4}h_{\mu \nu }h^{\mu \nu }\bar{R}+R^{(2)}\right) %
\biggr]\,,  \label{eq:3rdRphi2}
\end{equation}%
where we used the simplification that on a maximally symmetric background
one has that the first variation of $R^{(1)}$ is zero. The expressions for $%
R^{(2)}$ and $R^{(3)}$ are given in subsection \ref{sec_expand}. For the
case of $R^{2}$ piece of action the third variation is given by,
\begin{equation}
S_{R^{2}}^{(3)}=a\epsilon ^{3}\int \,\mathrm{d}^{d}x\sqrt{-\bar{g}}\biggl[%
\frac{1}{6}h^{\mu }{}_{\alpha }h^{\alpha }{}_{\beta }h^{\beta }{}_{\mu }\bar{%
R}+2\bar{R}R^{(3)}\biggr]\,.  \label{eq:3VarR2}
\end{equation}%
We notice that this expansion has some similar structure with the expansion
of $R\phi ^{2}$ piece. The third order expansion for the matter sector is
\begin{eqnarray}
&&S_{\mathrm{matter}}^{(3)}=\epsilon ^{3}\int \,\mathrm{d}^{d}x\sqrt{-\bar{g}%
}\biggl[\frac{1}{2}h^{\mu \nu }\partial _{\mu }\varphi \partial _{\nu
}\varphi -\frac{1}{6}V^{\prime \prime \prime }(\bar{\phi})\varphi ^{3}+\frac{%
1}{4}h^{\mu \nu }h_{\mu \nu }(m^{2}+V^{\prime }(\bar{\phi}))\varphi  \notag
\label{eq:3VarMat} \\
&&+\frac{1}{6}h^{\mu }{}_{\alpha }h^{\alpha }{}_{\beta }h^{\beta }{}_{\mu
}\left( \frac{1}{2}m^{2}\bar{\phi}^{2}+V(\bar{\phi})\right) \biggr]\,.
\end{eqnarray}%
We can combine all the pieces of the third order expansion of the action and
write it in a unified form. This is given by,
\begin{eqnarray}
&&S^{(3)}=\epsilon ^{3}\int \,\mathrm{d}^{d}x\sqrt{-\bar{g}}\biggl[\biggl\{%
\frac{(d+2)\xi \bar{R}\bar{\phi}^{2}}{12(d-1)}+\frac{(d+5)a\bar{R}^{2}}{%
6(d-1)}-\frac{1}{6}\left( \frac{1}{2}m^{2}\bar{\phi}^{2}+V(\bar{\phi}%
)\right) \biggr\}h^{\mu }{}_{\alpha }h^{\alpha }{}_{\beta }h^{\beta }{}_{\mu
}  \notag  \label{eq:3VarAct} \\
&&+\left( \frac{1}{2}\xi \bar{\phi}^{2}+2a\bar{R}\right) \biggl\{-(\bar{%
\nabla}_{\nu }h^{\lambda }{}_{\alpha })h_{\rho }{}^{\nu }(\bar{\nabla}%
_{\lambda }h^{\alpha \rho })+\frac{1}{2}(\bar{\nabla}^{\rho }h^{\lambda
\alpha })h_{\nu \lambda }(\bar{\nabla}_{\alpha }h_{\rho }{}^{\nu })+\frac{1}{%
2}h^{\nu \sigma }h^{\rho \alpha }\bar{\nabla}_{\rho }\bar{\nabla}_{\alpha
}h_{\nu \sigma }  \notag \\
&&+\frac{1}{4}(\bar{\nabla}_{\mu }h^{\rho \lambda })(\bar{\nabla}_{\nu
}h_{\rho \lambda })h^{\mu \nu }+\frac{1}{2}(\bar{\nabla}^{\rho }h^{\mu \nu
})h_{\mu }{}^{\lambda }(\bar{\nabla}_{\rho }h_{\lambda \nu })+\frac{1}{2}%
h^{\mu \nu }h_{\mu }{}^{\lambda }\Box h_{\lambda \nu }\biggr\}+\frac{1}{2}%
h^{\mu \nu }\partial _{\mu }\varphi \partial _{\nu }\varphi  \notag \\
&&-\frac{1}{6}V^{\prime \prime \prime }(\bar{\phi})\varphi ^{3}+\frac{1}{4}%
h^{\mu \nu }h_{\mu \nu }(m^{2}+V^{\prime }(\bar{\phi}))\varphi \biggr]\,.
\end{eqnarray}%
To obtain these one has to perform several integrations by parts which
allowed us to make use of the Landau gauge and trace-free condition. The
surface terms generated in this computation are included in appendix \ref%
{bound}.


\subsection{AdS/CFT discussion}

\label{sec_adsreln} 

After the scalar field receives the vev, the gravity action has an
Einstein-Hilbert term, as well as terms of matter fields and higher
derivative terms. The Einstein-Hilbert term is induced when the scalar
acquires a non-zero vacuum expectation value. At the same time, we obtained
the maximally symmetric AdS background and the dynamics of its fluctuation
fields, and we can use the AdS/CFT correspondence \cite%
{Maldacena:1997re,Gubser:1998bc,Witten:1998qj}. The dual field theory is in
its conformally invariant vacuum \cite{Osborn:1993cr}. The fluctuation
fields on the AdS background are obtained in the above subsections.

The fluctuation of the scalar field is $\varphi $. The quadratic part of the
Lagrangian of $\varphi $ is
\begin{equation}
L(\varphi )=-\frac{1}{2}(\partial \varphi )^{2}-\frac{1}{2}m_{\varphi
}^{2}\varphi ^{2},
\end{equation}%
where\vspace{1pt}%
\begin{equation}
m_{\varphi }^{2}=m^{2}-\xi \bar{R}+V^{\prime \prime }(\bar{\phi})={V}%
^{\prime \prime }{({\bar{\phi}})}-{V^{\prime }({\bar{\phi}})/{\bar{\phi}}},
\label{eq:mass_phi}
\end{equation}%
and we have used Eq. (\ref{bar_R_1}).

The operator dual to $\varphi $ is a scalar operator $O$ with scaling
dimension $\Delta =\Delta _{+}$ \cite{Witten:1998qj},\cite{Gubser:1998bc},
\begin{eqnarray}
\Delta _{\pm } &=&\frac{d-1}{2}\pm \sqrt{(\frac{d-1}{2})^{2}+m_{\varphi
}^{2}R_{AdS}^{2}}  \notag \\
&=&\frac{d-1}{2}\pm \sqrt{(\frac{d-1}{2})^{2}+\frac{{V^{\prime }({\bar{\phi}}%
)}-{V}^{\prime \prime }{({\bar{\phi}}){\bar{\phi}}}}{{V^{\prime }({\bar{\phi}%
})+m^{2}{\bar{\phi}}}}\xi d(d-1)},
\end{eqnarray}%
where we used Eq. (\ref{eq:mass_phi}). The Breitenlohner-Freedman (BF) bound
\cite{Breitenlohner:1982jf} demands that
\begin{equation}
\frac{1}{4}(d-1)+\frac{{V^{\prime }({\bar{\phi}})}-{V}^{\prime \prime }{({%
\bar{\phi}}){\bar{\phi}}}}{{V^{\prime }({\bar{\phi}})+m^{2}{\bar{\phi}}}}\xi
d\geqslant 0.
\end{equation}%
The corresponding two-point function of the scalar operator is%
\begin{equation}
\langle O(x)O(x^{\prime })\rangle =\frac{d_{O}}{|x-x^{\prime }|^{2\Delta }},
\end{equation}%
where $d_{O}$ is the normalisation factor. The three-point functions are
related to the cubic couplings in the third order variation of the action.

The AdS radius in the Planck unit is related to the number of degrees of
freedom of the fields in the field theory side. We can derive that the
central charge is
\begin{equation}
c_{T}\propto \left( \frac{R_{AdS}}{l_{p}}\right) ^{d-2}=\left( \frac{-\xi
d(d-1)}{{m^{2}+V^{\prime }({\bar{\phi}})/{\bar{\phi}}}}\right) ^{(d-2)/2}%
\left[ 8\pi (4a{\bar{R}}+\xi {\bar{\phi}}^{2})\right] .
\end{equation}

The AdS solution is perturbatively stable when the fluctuation modes have
mass-squared above the Breitenlohner-Freedman (BF) bound \cite%
{Breitenlohner:1982jf}. We find that the solutions are perturbatively
stable, with the condition above, in which the AdS background is stable
against small fluctuations.


\section{Conclusions and Discussions}

\label{sec_conc} 


In this paper, we analyzed a higher derivative gravity model coupled
non-minimally with scalar field with a general potential. The aim of the
paper is to see whether such an action can have AdS geometry as solution to
equation of motion. It has been seen that such an action can have an AdS
geometry for a wide class of potential for field $\phi $ and certain choice
of parameters. These actions have been widely studied in the context of
induced gravity and higher-derivative gravity in four or more general
spacetime dimensions, where the known low energy Einstein-Hilbert gravity
emerges from a UV well defined model of gravity. Then it is natural to
consider that such actions anticipate AdS geometry as solutions and also
their relation to the extensively studied AdS/CFT.

Here, we start by looking into equation of motions to look for AdS geometry.
We found the AdS solutions in various circumstances. We then compute the
second and third variation of the action. For this we compute first the
expansion of various geometrical quantities up to third order in the
fluctuation fields. These expansions are then used to compute the expansion
of the action of the full theory. We choose to work in Landau gauge where
the propagation of longitudinal modes is suppressed. Moreover, in flat
spacetime the usual harmonic gauge fixing condition doesn't fix the
invariance in the field $h_{\mu \nu }$ entirely. There is a residual freedom
left which is caused by dilatation invariance. This can be fixed by using
the trace-free condition. In AdS geometry with fixed AdS radius such
dilatation invariance is broken, which means one lacks the freedom to
have conformal transformation with arbitrary conformal factor. The
trace-free condition has also been used in the context of asymptotic safety.

The expansion of the geometrical quantities is used to obtain the expansion
of the action of the theory up to the third order. In holographic language
the second order terms are used to obtain mass-parameter for various field
modes, while the third order terms lead to various correlators on the
boundary. More details of the study of these relations deserve further
investigation and we leave them for future exploration.

We obtain a new scenario for AdS solutions, in particular in the context of
induced gravity. This scenario can have various generalizations. We can
couple other fields with the system. For example, we can add vector fields
or fermions into the system. This scenario potentially has interesting
applications for holographic correspondence. On the other hand, the modified
gravity theory and the Einstein gravity have slightly different physical
observables. For example, the wave form of the gravitational waves predicted
from scalar-tensor theory and the Einstein gravity are slightly different.
Because the physical observables are slightly variant, hence understanding
the detailed dictionary of holography for modified gravity theory is
interesting and deserves future investigations.


\bigskip \centerline{\bf Acknowledgements}

We would like to thank Y. An, J. Maldacena, A. Petkou, J. Simon, W. Song, N.
Su, H. Wang, and S.-T. Yau for communications or discussions. The work was
supported in part by Yau Mathematical Sciences Center and by Tsinghua
University. GN would like to thank Prof. Tianjun Li for support and
encouragement during the course of this work.


\appendix
\section{Boundary terms}
\label{bound} 

Here in this appendix we write the boundary terms that are generated while
performing integration by parts during the course of evaluation at various
orders. These may play some role in CFT on the boundary.

At second order the boundary terms are given by,
\begin{eqnarray}
&&(\partial \mathcal{M})^{(2)}=\epsilon ^{2}\left( \frac{1}{2}\xi \bar{\phi}%
^{2}+2a\bar{R}\right) \int \mathrm{d}^{d}x\sqrt{-\bar{g}}\biggl[-\frac{1}{2}%
\bar{\nabla}_{\nu }(h^{\nu \sigma }\bar{\nabla}_{\rho }h^{\rho }{}_{\sigma
})+\frac{1}{2}\bar{\nabla}_{\nu }(h^{\nu \sigma }\bar{\nabla}_{\sigma }h)-%
\frac{1}{4}\bar{\nabla}_{\lambda }(h\bar{\nabla}^{\lambda }h)  \notag
\label{eq:2ndB} \\
&&+\frac{1}{4}\bar{\nabla}_{\rho }(h^{\nu \lambda }\bar{\nabla}^{\rho
}h_{\nu \lambda })-\frac{1}{2}\bar{\nabla}_{\lambda }(h^{\rho \nu }\bar{%
\nabla}_{\rho }h^{\lambda }{}_{\nu })+\bar{\nabla}_{\rho }(\bar{g}^{\nu
\sigma }\Gamma _{\nu }^{(2)}{}^{\rho }{}_{\sigma })-\bar{\nabla}_{\nu }(\bar{%
g}^{\nu \sigma }\Gamma _{\rho }^{(2)}{}^{\rho }{}_{\sigma })+\frac{1}{2}\bar{%
\nabla}_{\alpha }(\bar{g}^{\alpha \beta }\varphi \bar{\nabla}_{\beta
}\varphi )\biggr]\,.  \notag \\
&&
\end{eqnarray}%
At the third order the boundary terms are more complicated, as one has to
perform more number of integrations by parts. And we have
\begin{eqnarray}
&&(\partial \mathcal{M})^{(3)}=\epsilon ^{3}\left( \frac{1}{2}\xi \bar{\phi}%
^{2}+2a\bar{R}\right) \int \mathrm{d}^{d}x\sqrt{-\bar{g}}\biggl[\bar{\nabla}%
_{\rho }(\bar{g}^{\nu \sigma }\Gamma _{\nu }^{(3)}{}^{\rho }{}_{\sigma })-%
\bar{\nabla}_{\nu }(\bar{g}^{\nu \sigma }\Gamma _{\rho }^{(3)}{}^{\rho
}{}_{\sigma })-\frac{1}{2}\bar{\nabla}_{\nu }(h^{\lambda \alpha }h^{\rho
}{}_{\lambda }\bar{\nabla}^{\nu }h_{\alpha \rho })  \notag  \label{eq:3rdB}
\\
&&+\frac{1}{2}\bar{\nabla}_{\nu }(h^{\lambda \alpha }h^{\rho \nu }\bar{\nabla%
}_{\lambda }h_{\alpha \rho })-\frac{1}{2}\bar{\nabla}_{\lambda }(h^{\lambda
\alpha }h^{\rho }{}_{\nu }\bar{\nabla}^{\nu }h_{\alpha \rho })+\frac{1}{2}%
\bar{\nabla}_{\rho }(h^{\lambda \alpha }h_{\nu \lambda }\bar{\nabla}^{\rho
}h^{\nu }{}_{\alpha })-\frac{1}{2}\bar{\nabla}_{\rho }(h^{\lambda \alpha
}h^{\nu }{}_{\lambda }\bar{\nabla}_{\alpha }h^{\rho }{}_{\nu })  \notag \\
&&+\frac{1}{2}\bar{\nabla}_{\nu }(h^{\nu \sigma }\Gamma _{\rho
}^{(2)}{}^{\rho }{}_{\sigma })-\frac{1}{4}\bar{\nabla}_{\nu }(h^{\mu \nu
}h_{\rho \lambda }\bar{\nabla}_{\mu }h^{\rho \lambda })+\frac{1}{2}\bar{%
\nabla}_{\lambda }(h^{\mu }{}_{\nu }h^{\rho }{}_{\mu }\bar{\nabla}_{\rho
}h^{\lambda \nu })-\frac{1}{2}\bar{\nabla}_{\lambda }(h^{\mu }{}_{\nu
}h{}_{\rho \mu }\bar{\nabla}^{\lambda }h^{\rho \nu })\biggr]  \notag \\
&&+\frac{\epsilon ^{3}}{2}\xi \bar{\phi}\int \mathrm{d}^{d}x\sqrt{-\bar{g}}%
\biggl[\bar{\nabla}_{\rho }(\varphi \bar{g}^{\nu \sigma }\Gamma _{\nu
}^{(2)}{}^{\rho }{}_{\sigma })-\bar{\nabla}_{\nu }(\varphi \bar{g}^{\nu
\sigma }\Gamma _{\rho }^{(2)}{}^{\rho }{}_{\sigma })+\frac{1}{2}\bar{\nabla}%
_{\lambda }(\varphi h^{\mu \nu }\bar{\nabla}^{\lambda }h_{\mu \nu })+\frac{1%
}{4}\bar{\nabla}_{\lambda }((\bar{\nabla}^{\lambda }\varphi )h^{\mu \nu
}h_{\mu \nu })  \notag \\
&&-\bar{\nabla}_{\lambda }((\bar{\nabla}^{\rho }\varphi )h_{\rho \nu
}h^{\lambda \nu })-\bar{\nabla}_{\rho }(\varphi h_{\lambda \nu }\bar{\nabla}%
^{\lambda }h^{\rho \nu })\biggr]\,.
\end{eqnarray}



\begin{thebibliography}{0}%
\makeatletter
\providecommand \@ifxundefined [1]{%
 \@ifx{#1\undefined}
}%
\providecommand \@ifnum [1]{%
 \ifnum #1\expandafter \@firstoftwo
 \else \expandafter \@secondoftwo
 \fi
}%
\providecommand \@ifx [1]{%
 \ifx #1\expandafter \@firstoftwo
 \else \expandafter \@secondoftwo
 \fi
}%
\providecommand \natexlab [1]{#1}%
\providecommand \enquote  [1]{``#1''}%
\providecommand \bibnamefont  [1]{#1}%
\providecommand \bibfnamefont [1]{#1}%
\providecommand \citenamefont [1]{#1}%
\providecommand \href@noop [0]{\@secondoftwo}%
\providecommand \href [0]{\begingroup \@sanitize@url \@href}%
\providecommand \@href[1]{\@@startlink{#1}\@@href}%
\providecommand \@@href[1]{\endgroup#1\@@endlink}%
\providecommand \@sanitize@url [0]{\catcode `\\12\catcode `\$12\catcode
  `\&12\catcode `\#12\catcode `\^12\catcode `\_12\catcode `\%12\relax}%
\providecommand \@@startlink[1]{}%
\providecommand \@@endlink[0]{}%
\providecommand \url  [0]{\begingroup\@sanitize@url \@url }%
\providecommand \@url [1]{\endgroup\@href {#1}{\urlprefix }}%
\providecommand \urlprefix  [0]{URL }%
\providecommand \Eprint [0]{\href }%
\providecommand \doibase [0]{http://dx.doi.org/}%
\providecommand \selectlanguage [0]{\@gobble}%
\providecommand \bibinfo  [0]{\@secondoftwo}%
\providecommand \bibfield  [0]{\@secondoftwo}%
\providecommand \translation [1]{[#1]}%
\providecommand \BibitemOpen [0]{}%
\providecommand \bibitemStop [0]{}%
\providecommand \bibitemNoStop [0]{.\EOS\space}%
\providecommand \EOS [0]{\spacefactor3000\relax}%
\providecommand \BibitemShut  [1]{\csname bibitem#1\endcsname}%
\let\auto@bib@innerbib\@empty
\end{thebibliography}%


\begin{thebibliography}{99}

\bibitem{tHooft:1974toh} G.~'t Hooft and M.~J.~G.~Veltman, ``One loop
divergencies in the theory of gravitation,'' Ann.\ Inst.\ H.\ Poincare
Phys.\ Theor.\ A \textbf{20} (1974) 69.

\bibitem{Goroff:1985sz} M.~H.~Goroff and A.~Sagnotti, ``Quantum Gravity At
Two Loops,'' Phys.\ Lett.\ \textbf{160B} (1985) 81.

\bibitem{Goroff:1985th} M.~H.~Goroff and A.~Sagnotti, ``The Ultraviolet
Behavior of Einstein Gravity,'' Nucl.\ Phys.\ B \textbf{266} (1986) 709.

\bibitem{Weinberg:1976xy} S.~Weinberg, ``Critical Phenomena for Field
Theorists,'' HUTP-76-160. 

\bibitem{Weinberg:1980gg} S.~Weinberg, ``Ultraviolet Divergences In Quantum
Theories Of Gravitation,'' In General Relativity: An Einstein centenary
survey, ed. S. W. Hawking and W. Israel, Cambridge University Press (1979)
790-831.

\bibitem{Percacci:2007sz} R.~Percacci, ``Asymptotic Safety,'' In Oriti, D. :
Approaches to quantum gravity, 111-128 [arXiv:0709.3851 [hep-th]].

\bibitem{Stelle:1976gc} K.~S.~Stelle, ``Renormalization of Higher Derivative
Quantum Gravity,'' Phys.\ Rev.\ D \textbf{16} (1977) 953.

\bibitem{Salam:1978fd} A.~Salam and J.~A.~Strathdee, ``Remarks on
High-energy Stability and Renormalizability of Gravity Theory,'' Phys.\
Rev.\ D \textbf{18}, 4480 (1978). 

\bibitem{Julve:1978xn} J.~Julve and M.~Tonin, ``Quantum Gravity with Higher
Derivative Terms,'' Nuovo Cim.\ B \textbf{46}, 137 (1978).

\bibitem{Narain:2011gs} G.~Narain and R.~Anishetty, ``Short Distance Freedom
of Quantum Gravity,'' Phys.\ Lett.\ B \textbf{711}, 128 (2012)
[arXiv:1109.3981 [hep-th]].

\bibitem{Narain:2012nf} G.~Narain and R.~Anishetty, ``Unitary and
Renormalizable Theory of Higher Derivative Gravity,'' J.\ Phys.\ Conf.\
Ser.\ \textbf{405}, 012024 (2012) 
[arXiv:1210.0513 [hep-th]].

\bibitem{Fradkin:1981hx} E.~S.~Fradkin and A.~A.~Tseytlin, ``Renormalizable
Asymptotically Free Quantum Theory of Gravity,'' Phys.\ Lett.\ \textbf{104B}%
, 377 (1981). 

\bibitem{Fradkin:1981iu} E.~S.~Fradkin and A.~A.~Tseytlin, ``Renormalizable
asymptotically free quantum theory of gravity,'' Nucl.\ Phys.\ B \textbf{201}
(1982) 469. 

\bibitem{Alvarez-Gaume:2015rwa} L.~Alvarez-Gaume, A.~Kehagias, C.~Kounnas,
D.~Lust and A.~Riotto, ``Aspects of Quadratic Gravity,'' Fortsch.\ Phys.\
\textbf{64}, no. 2-3, 176 (2016) 
[arXiv:1505.07657 [hep-th]]. 

\bibitem{Barth:1983hb} N.~H.~Barth and S.~M.~Christensen, ``Quantizing
Fourth Order Gravity Theories. 1. The Functional Integral,'' Phys.\ Rev.\ D
\textbf{28}, 1876 (1983). 

\bibitem{Avramidi:1985ki} I.~G.~Avramidi and A.~O.~Barvinsky, ``Asymptotic
Freedom In Higher Derivative Quantum Gravity,'' Phys.\ Lett.\ \textbf{159B},
269 (1985). 

\bibitem{Buchbinder:1989jd} I.~L.~Buchbinder, O.~K.~Kalashnikov,
I.~L.~Shapiro, V.~B.~Vologodsky and J.~J.~Wolfengaut, ``The Stability of
Asymptotic Freedom in Grand Unified Models Coupled to $R^{2}$ Gravity,''
Phys.\ Lett.\ B \textbf{216}, 127 (1989). 

\bibitem{Shapiro:1989dq} I.~L.~Shapiro, ``Asymptotic Behavior of Effective
Yukawa Coupling Constants in Quantum $R^{2}$ Gravity With Matter,'' Class.\
Quant.\ Grav.\ \textbf{6}, 1197 (1989). 

\bibitem{Odintsov:1989gz} S.~D.~Odintsov, ``The Parametrization Invariant
and Gauge Invariant Effective Actions in Quantum Field Theory,'' Fortsch.\
Phys.\ \textbf{38}, 371 (1990). 

\bibitem{Zee:1978wi} A.~Zee, ``A Broken Symmetric Theory of Gravity,''
Phys.\ Rev.\ Lett.\ \textbf{42} (1979) 417. 

\bibitem{Zee:1980sj} A.~Zee, ``Spontaneously Generated Gravity,'' Phys.\
Rev.\ D \textbf{23} (1981) 858. 

\bibitem{Adler:1980pg} S.~L.~Adler, ``A Formula for the Induced
Gravitational Constant,'' Phys.\ Lett.\ \textbf{95B} (1980) 241.

\bibitem{Adler:1980bx} S.~L.~Adler, ``Order R Vacuum Action Functional in
Scalar Free Unified Theories with Spontaneous Scale Breaking,'' Phys.\ Rev.\
Lett.\ \textbf{44} (1980) 1567. 

\bibitem{Adler:1980md} S.~L.~Adler, ``Induced gravitation,'' AIP Conf.\
Proc.\ \textbf{68} (1980) 915. 

\bibitem{Adler:1982ri} S.~L.~Adler, \textquotedblleft Einstein Gravity as a
Symmetry Breaking Effect in Quantum Field Theory,\textquotedblright\ Rev.\
Mod.\ Phys.\ \textbf{54} (1982) 729.

\bibitem{Zeldovich1:967gd} Y.~B.~Zeldovich, \textquotedblleft Cosmological
Constant and Elementary Particles,\textquotedblright\ JETP Lett.\ \textbf{6}
(1967) 316.

\bibitem{Sakharov:1967pk} A.~D.~Sakharov, \textquotedblleft Vacuum quantum
fluctuations in curved space and the theory of
gravitation,\textquotedblright\ Sov.\ Phys.\ Dokl.\ \textbf{12}, 1040 (1968)
[Sov.\ Phys.\ Usp.\ \textbf{34}, 394 (1991)] [Gen.\ Rel.\ Grav.\ \textbf{32}%
, 365 (2000)]. 

\bibitem{Chudnovsky:1976zj} E.~M.~Chudnovsky, ``The Spontaneous Conformal
Symmetry Breaking and Higgs Model,'' Theor.\ Math.\ Phys.\ \textbf{35}, 538
(1978) [Teor.\ Mat.\ Fiz.\ \textbf{35}, 398 (1978)].

\bibitem{Nepomechie:1983yq} R.~I.~Nepomechie, ``Einstein Gravity as the
Low-energy Effective Theory of Weyl Gravity,'' Phys.\ Lett.\ \textbf{136B},
33 (1984). 

\bibitem{Zee:1983mj} A.~Zee, ``Einstein Gravity Emerging From Quantum Weyl
Gravity,'' Annals Phys.\ \textbf{151}, 431 (1983).

\bibitem{Salvio:2016vxi} A.~Salvio, ``Solving the Standard Model Problems in
Softened Gravity,'' Phys.\ Rev.\ D \textbf{94}, no. 9, 096007 (2016)
[arXiv:1608.01194 [hep-ph]]. 

\bibitem{Einhorn:2014gfa} M.~B.~Einhorn and D.~R.~T.~Jones, ``Naturalness
and Dimensional Transmutation in Classically Scale-Invariant Gravity,'' JHEP
\textbf{1503}, 047 (2015) 
[arXiv:1410.8513[hep-th]]. 

\bibitem{Salvio:2014soa} A.~Salvio and A.~Strumia, ``Agravity,'' JHEP
\textbf{1406} (2014) 080 
[arXiv:1403.4226[hep-ph]]. 

\bibitem{Jones:2015son} T.~Jones and M.~Einhorn, ``Quantum Gravity and
Dimensional Transmutation,'' PoS PLANCK \textbf{2015}, 061 (2015).

\bibitem{Einhorn:2015lzy} M.~B.~Einhorn and D.~R.~T.~Jones, ``Induced
Gravity I: Real Scalar Field,'' JHEP \textbf{1601}, 019 (2016)
[arXiv:1511.01481 [hep-th]]. 

\bibitem{Einhorn:2016mws} M.~B.~Einhorn and D.~R.~T.~Jones, ``Induced
Gravity II: Grand Unification,'' JHEP \textbf{1605}, 185 (2016)
[arXiv:1602.06290 [hep-th]]. 

\bibitem{Cooper:1982du} F.~Cooper and G.~Venturi, ``Cosmology and Broken
Scale Invariance,'' Phys.\ Rev.\ D \textbf{24}, 3338 (1981).

\bibitem{Finelli:2007wb} F.~Finelli, A.~Tronconi and G.~Venturi, ``Dark
Energy, Induced Gravity and Broken Scale Invariance,'' Phys.\ Lett.\ B
\textbf{659}, 466 (2008) 
[arXiv:0710.2741[astro-ph]].

\bibitem{Maldacena:1997re} J.~M.~Maldacena, ``The Large N limit of
superconformal field theories and supergravity,'' Adv.\ Theor.\ Math.\
Phys.\ \textbf{2} (1998) 231
[hep-th/9711200].

\bibitem{Gubser:1998bc} S.~S.~Gubser, I.~R.~Klebanov and A.~M.~Polyakov,
``Gauge theory correlators from noncritical string theory,'' Phys.\ Lett.\ B
\textbf{428} (1998) 105 
[hep-th/9802109]. 

\bibitem{Witten:1998qj} E.~Witten, ``Anti-de Sitter space and holography,''
Adv.\ Theor.\ Math.\ Phys.\ \textbf{2} (1998) 253
[hep-th/9802150]. 

\bibitem{Rangamani:2016dms} M.~Rangamani and T.~Takayanagi, ``Holographic
Entanglement Entropy,'' Lect.\ Notes Phys.\ \textbf{931} (2017) 1
[arXiv:1609.01287 [hep-th]].

\bibitem{VanRaamsdonk:2010pw} M.~Van Raamsdonk, ``Building up spacetime with
quantum entanglement,'' Gen.\ Rel.\ Grav.\ \textbf{42} (2010) 2323 [Int.\
J.\ Mod.\ Phys.\ D \textbf{19} (2010) 2429]
[arXiv:1005.3035 [hep-th]].

\bibitem{Horowitz:2006ct} G.~T.~Horowitz and J.~Polchinski,
\textquotedblleft Gauge/gravity duality,\textquotedblright\ In: D. Oriti,
Approaches to quantum gravity, Cambridge University Press [gr-qc/0602037].

\bibitem{Koch:2009gq} R.~de Mello Koch and J.~Murugan, ``Emergent
Spacetime,'' In: Foundations of Space and Time, Cambridge University Press
[arXiv:0911.4817 [hep-th]]. 

\bibitem{Schoen:1984} R. Schoen, ``Conformal deformation of a Riemannian
metric to constant scalar curvature,'' J. Differential Geom. \textbf{20}
(1984), no. 2, 479--495.

\bibitem{Lee:1987} J. M. Lee, T. H. Parker, ``The Yamabe problem,'' Bull.
Amer. Math. Soc. \textbf{17} (1987), no. 1, 37--91.

\bibitem{DeWitt3} B.~S.~DeWitt, ``A Gauge Invariant Effective Action,'' In:
Oxford 1980, Proceedings, Quantum Gravity 2, 449-487.

\bibitem{Abbott} L.~F.~Abbott, ``The Background Field Method Beyond One
Loop,'' Nucl.\ Phys.\ B \textbf{185}, 189 (1981).

\bibitem{Faddeev:1967fc} L.~D.~Faddeev and V.~N.~Popov, ``Feynman Diagrams
for the Yang-Mills Field,'' Phys.\ Lett.\ \textbf{25B} (1967) 29.

\bibitem{DeWitt:1964yg} B.~S.~DeWitt, ``Theory of radiative corrections for
non-abelian gauge fields,'' Phys.\ Rev.\ Lett.\ \textbf{12} (1964) 742.

\bibitem{Antoniadis:1985ub} I.~Antoniadis, J.~Iliopoulos and T.~N.~Tomaras,
``Gauge Invariance in Quantum Gravity,'' Nucl.\ Phys.\ B \textbf{267} (1986)
497. 

\bibitem{Ohta:2015zwa} N.~Ohta and R.~Percacci, ``Ultraviolet Fixed Points
in Conformal Gravity and General Quadratic Theories,'' Class.\ Quant.\
Grav.\ \textbf{33} (2016) 035001 
[arXiv:1506.05526 [hep-th]].

\bibitem{Breitenlohner:1982jf} P.~Breitenlohner and D.~Z.~Freedman,
``Stability in Gauged Extended Supergravity,'' Annals Phys.\ \textbf{144}
(1982) 249. 

\bibitem{Osborn:1993cr} H.~Osborn and A.~C.~Petkou, ``Implications of
conformal invariance in field theories for general dimensions,'' Annals
Phys.\ \textbf{231} (1994) 311 
[hep-th/9307010]. 
\end{thebibliography}
\end{document}